\documentclass[preprint,pra,aps]{revtex4}

\usepackage{epsfig}

\begin{document}

\author{David P. DiVincenzo$^1$ and Daniel Loss$^2$}

\title{Rigorous Born Approximation and beyond for the Spin-Boson Model}

\affiliation{\vspace*{1.2ex}$^1$
            \hspace*{0.5ex}{IBM Research Division, T. J. Watson Research Center,
P.O. Box 218, Yorktown Heights, NY 10598, USA}}
\affiliation{\vspace*{1.2ex}$^2$
            \hspace*{0.5ex}{Department of Physics and Astronomy, University of Basel,
            Klingelbergstrasse 82, CH-4056 Basel, Switzerland}}

\date{\today}

\begin{abstract}

Within the lowest-order Born approximation, we present an exact
calculation of the time dynamics of the spin-boson model in the
ohmic regime.  We observe non-Markovian effects at zero
temperature that scale with the system-bath coupling strength and
cause qualitative changes in the evolution of coherence at
intermediate times of order of the oscillation period.  These
changes could significantly affect the performance of these
systems as qubits. In the biased case, we find a prompt loss of
coherence at these intermediate times, whose decay rate is set by
$\sqrt{\alpha}$, where $\alpha$ is the coupling strength to the
environment.  We also explore the calculation of the next order
Born approximation: we show that, at the expense of very large
computational complexity, interesting physical quantities can be
rigorously computed at fourth order using computer algebra,
presented completely in an accompanying Mathematica file. We
compute the $O(\alpha)$ corrections to the long time behavior of
the system density matrix; the result is identical to the reduced
density matrix of the equilibrium state to the same order in
$\alpha$. All these calculations indicate precision experimental
tests that could confirm or refute the validity of the spin-boson
model in a variety of systems.


\end{abstract}

\maketitle

\section{Introduction}

Novel solid state devices that can control spin degrees of freedom
of individual electrons\cite{LK,DL97}, or discrete quantum states
in superconducting circuits\cite{Naka,delft,vion,RMPkarl}, show
promise in realizing the ideal of the completely controllable
two-state quantum system, weakly coupled to its environment, that
is the essential starting point for qubit operation in quantum
computation.  From a fundamental point of view, these experimental
successes also bring us close to embodying the ideal test of
quantum coherence as envisioned by Leggett many years
ago\cite{Leggett}, in which a simple quantum system is placed in a
known initial state, is allowed to evolve for a definite time $t$
under the action of its own Hamiltonian and under the influence of
decoherence from the environment, and is then measured.

Recent experiments, starting with \cite{Naka}, show that this
ideal test can be implemented in practice.  The decay of quantum
oscillations due to environmental decoherence is
now\cite{Naka,delft,vion,RMPkarl} sufficiently weak that some tens
of coherent oscillations can be observed. If quantum computation
is to become a reality, it is believed\cite{DL} that these systems
will eventually need to achieve even lower levels of decoherence,
such that thousands or tens of thousands of coherent oscillations
could be observed.  This prospect of producing experiments with
ultra-long coherence times in quantum two state systems offers a
new challenge for theoretical modelling of decoherence.  Despite
the many years of work\cite{weiss,Haenggi} following on Leggett's
initial proposals, there has never been a full, systematic
analysis of the most popular description of these systems, the
spin-boson model, in the limit of very weak coupling to the
environment.

In this paper we provide an exact analysis of the weak coupling
limit of the spin boson model for the ohmic heat bath, and in the
low temperature limit.  In this limit the Born approximation (to
the self energy) should become essentially exact, and we make no
other approximations in our solutions --- in particular, no Markov
approximation is made. As other workers have recently
emphasized\cite{braun,privman}, understanding the details of the
short-time dynamics of this model is especially crucial for the
operation of these systems as qubits.

We find important, new, non-Markovian effects in this regime.  At
lowest order in the Born expansion of the self energy
superoperator, the time dynamics of the model rigorously separates
into a sum of strictly exponential pieces (the usual ``$T_1$" and
``$T_2$" decays of the Bloch-Redfield model) plus two distinct
non-exponential pieces that arise, technically speaking, from two
different kinds of branch cuts in the Laplace transform of the
solution of the generalized master equation that we obtain.

These two contributions both have power-law forms at long times,
$t>T_1,T_2$, and thus formally dominate the exponentially-decaying
parts.  But more interesting is that they both give new structure
to the time evolution at intermediate times $t$, $1/\omega_c <t<
T_1,T_2$; this structure typically occurs for $t$ on the order of
the oscillation period. (Here, $\omega_c$ is a high frequency
cut-off of the bath modes, defining the very short time regime,
$t<1/\omega_c$, which is of no interest here.)  We can explain our
results in the language of the double-well potential, where the
two quantum states are ``left" and ``right" ($L$/$R$), the $t=0$
state is pure $L$, and the system oscillates in time via tunneling
from $L$ to $R$.  The first branch-cut contribution is most
important in the unbiased case ($L$ and $R$ energies degenerate)
and it causes the system, starting immediately in the first
quantum oscillation, to spend more time in the $R$ well, that is,
the {\em opposite} well from the one the system is in initially.
The second branch-cut contribution, present when the system is
biased, adds to the amplitude of the coherent oscillation, but
dies out after an intermediate time which scales like the inverse
square root of the interaction strength $\alpha$ with the bath.
This {\em prompt loss of coherence}, whose amplitude is
proportional to $\alpha$, changes qualitatively the picture of the
initial decay of coherence that is so important for discussions of
fault-tolerant quantum computation\cite{Preskill}.

Finally, we set up the next-order Born approximation and do some
initial calculations with it.  This involves computing the
self-energy of the master equation to fourth order in the
system-bath coupling. At this order the self energy is a sum of
thousands of separate terms; but we find that it is feasible to
compute various quantities of physical interest with the aid of
Mathematica.  As an illustration, we provide a full calculation of
the steady state system density matrix to order $\alpha$ in the
limit of low temperature, which requires the fourth-order self
energy. Given the enormous complexity of the calculation, we find
a very simple result for the corrections to steady state; they
turn out to be identical to those for the thermodynamic
equilibrium state calculated to the same order in $\alpha$.  Thus,
we are able to establish rigorously a very strong form of
ergodicity for the spin boson model at this order.

\section{Generalized Master Equation}

We are interested in studying the time dependence of the system
density matrix $\rho_S(t)={\rm Tr}_B \rho(t)$ with a
time-independent system Hamiltonian, and in the presence of a
fixed coupling to an environment.  An exact equation for $\rho_S$
-- the generalized master equation (GME) -- is\cite{FS}
\begin{eqnarray}
\dot\rho_S(t)&=&-iL_S\rho_S(t)-i\int_0^tdt'\Sigma_S(t-t')\rho_S(t'),
\label{p1_3e1_11_1}\\
\Sigma_S(t)&=&-i{\rm Tr}_BL_{SB}e^{-iQLt}L_{SB}\rho_B.
\label{p1_3e1_11_2}
\end{eqnarray}
Here the kernel $\Sigma_S(t)$ is the self energy superoperator,
the system-bath Hamiltonian is written
\begin{equation}
H=H_S+H_{SB}+H_B, \label{p2_11}
\end{equation}
(S = system, B = bath), the Liouvillian superoperator is defined
by $L_x\rho=[H_x,\rho]$, $\rho_B=e^{-\beta H_B}/Z$,
$\beta=1/k_BT$, $T$ is the temperature, and $Q$ is the projection
superoperator $Q=1-\rho_B{\rm Tr}_B$.
Eq. (\ref{p1_3e1_11_1}) is written for the case ${\rm
Tr}_BH_{SB}\rho_B=0$, and the total initial state is taken to be
of the form $\rho(0)=\rho_S(0)\otimes\rho_B$, for an arbitrary
$\rho_S(0)$.  Since we are interested in the case of weak coupling
to the bath, we will consider a systematic expansion in powers of
this coupling $L_{SB}$ in the self-energy operator $\Sigma_S(t)$.

Retention of only the lowest order term in this expansion, giving
the first Born approximation, is obtained\cite{CL} by the
replacement $e^{-iQLt}\rightarrow e^{-iQ(L_S+L_B)t}$
in Eq. (\ref{p1_3e1_11_2}).  Thus, in the lowest Born
approximation, the self energy becomes
\begin{equation}
\Sigma^{(2)}_S(t)=-i{\rm Tr}_BL_{SB}e^{-i(L_S+L_B)t}L_{SB}\rho_B.
\end{equation}
We have used the fact here that the expression is unaffected if
the $Q$ superoperator is dropped in the exponential.

We now proceed to solve the GME {\em with no further
approximations.}\cite{Herb}  This distinguishes our work from
previous efforts, in which various other approximations (secular,
rotating wave, Markov, ``non-interacting blips", short time) are
made (see, e.g., \cite{Leggett,Haenggi,weiss,braun,privman}). We
will find that, in particular, avoidance of the Markov
approximation endows the solution with qualitatively new features.

We will work out all our results for the ohmic spin-boson model,
for which the Hamiltonian is
\begin{eqnarray}
H_S&=&{\Delta\over 2}\sigma_x+{\epsilon\over 2}\sigma_z,\label{hss}\\
H_{SB}&=&\sigma_z\otimes[\sum_n c_n(b_n^\dagger+b_n)],\\
H_B&=&\sum_n\omega_nb_n^\dagger b_n.
\end{eqnarray}
Here $\sigma_{x,y,z}$ are the Pauli operators; we will use
$\sigma_0=I$ (identity operator).  Also, $b_n^\dagger$ and $b_n$
are the creation and annihilation operators of harmonic oscillator
$n$ of the bath.  With the spectral density defined as
\begin{equation}
J(\omega)\equiv\sum_nc_n^2\delta(\omega-\omega_n),
\end{equation}
the ``ohmic" case is defined by choosing the coefficients $c_n$
and the oscillator frequencies $\omega_n$ such that, in the limit
of a continuous spectrum,
\begin{equation}
J(\omega)={\alpha\over 2}\omega e^{-\omega/\omega_c}\label{ohm1}
\end{equation}
Here $\omega_c$ is an ultraviolet cutoff frequency.

The first few steps of the solution of the GME do not depend on
the details of this model; we need only assume that the system
Hilbert space is two dimensional, and the system-bath coupling has
the bilinear form, $H_{SB}=S\otimes X$
($S$ ($X$) is an operator in the system (bath) space).  Under
these general circumstances the GME (\ref{p1_3e1_11_1}) in the
Born approximation can be rewritten in an ordinary operator form:
\begin{eqnarray}
\langle\dot\sigma_\mu(t)\rangle&=&-i{\rm Tr}_S\sigma_
\mu[H_S,\rho_S(t)]-\int_0^tdt'I_\mu(t,t')\, ,\label{p2_2e2_61}\\
I_\mu(t,t')&=&I_{\mu
0}(t')+\sum_{\nu=1}^3I_{\mu\nu}(t')\langle\sigma_\nu(t-t')\rangle\, ,\label{p2_2e2_62}\\
I_{\mu\nu}(t')&=&{\rm Re}~\{C(-t'){\rm
Tr}_S\sigma_\nu(-t')[\sigma_\mu,S]S(-t')\}\, .\label{p2_2e2_63}
\end{eqnarray}
Here $\langle x\rangle\equiv{\rm Tr}_S x\rho_S$, and the bath
correlation function is
\begin{equation}
C(t)\equiv {\rm Tr}_B[XX(t)\rho_B]=C'(t)+iC''(t).\label{pII7e56}
\end{equation}
$C'$ and $C''$ denote the real and imaginary parts of the bath
correlator), and, for the spin-boson model, $X=\sum_n
c_n(b_n^\dagger+b_n)$.  The time dependent operators are in the
interaction picture, i.e.,
\begin{equation}
\Xi(t)=e^{i(H_S+H_B)t}\Xi e^{-i(H_S+H_B)t}, \label{pII6e352}
\end{equation}
for any operator $\Xi$.\footnote{Note that once the choice
$S=\sigma_z$ is made for the system part of the system-bath
interaction Hamiltonian, the system Hamiltonian can without loss
of generality be chosen to be of the form of Eq.
(\protect\ref{hss}) with some choice of the parameters $\Delta$
and $\epsilon$.}

The GME in Eq. (\ref{p2_2e2_61}) can be written in the matrix form
\begin{equation}
\langle\underline{\dot\sigma}(t)\rangle=
R*\langle\underline{\sigma}\rangle+\underline{k}.
\label{p2_10e2_49}
\end{equation}
Here $\underline{\sigma}$ denotes the vector
$(\sigma_x,\sigma_y,\sigma_z)^T$ and convolution is denoted
$A*B\equiv\int_0^tdt'A(t')B(t-t')$.  When the the system
Hamiltonian is chosen as in Eq. (\ref{hss}), and the system part
of the system-bath interaction Hamiltonian is $S=\sigma_z$, then
we have\footnote{Note that in this theory there is a very simple
relationship between $\sigma_y$ and $\sigma_z$:
$\Delta\sigma_y={\dot\sigma}_z$.}

\begin{eqnarray}
R(t)&=&\left(\begin{array}{ccc}-{E^2\over\Delta^2}\Gamma_1(t)&
-\epsilon\delta(t)+{E\over\Delta}K^+_y(t)&0\\
\epsilon\delta(t)-{E\over\Delta}K^+_y(t)&-\Gamma_y(t)&-\Delta\delta(t)\\
0&\Delta\delta(t)&0\end{array}\right),\label{p2_10e2_51}\\
\underline{k}(t)&=&\left(\begin{array}{ccc}-{E\over\Delta}k^-(t),&
-k_y^-(t),&0\end{array}\right)^T,\label{p2_10e2_492}
\end{eqnarray}
with
\begin{equation}
E=\sqrt{\epsilon^2+\Delta^2}.
\end{equation}
We have introduced the functions
\begin{eqnarray}
\Gamma_1(t)&=&{4\Delta^2\over
E^2}\cos(Et)C'(t),\label{p2_11e2_551}\\
\Gamma_y(t)&=&{4\Delta^2\over
E^2}\left (1+{\epsilon^2\over\Delta^2}\cos(Et)\right )C'(t),\label{p2_11e2_552}\\
K_y^+(t)&=&{4\epsilon\Delta\over
E^2}\sin(Et)C'(t),\label{p2_11e2_553}\\
k^-(t)&=&{4\Delta^2\over
E^2}\int_0^tdt'\sin(Et')C''(t'),\label{p2_11e2_554}\\
k_y^-(t)&=&{4\epsilon\Delta\over
E^2}\int_0^tdt'(1-\cos(Et'))C''(t').\label{p2_11e2_555}
\end{eqnarray}

Eq. (\ref{p2_10e2_49}) can be solved in the Laplace domain.
Defining the Laplace transform as
\begin{equation}
f(s)=\int_0^\infty e^{-st}f(t)dt, \label{p10_1e114}
\end{equation}
the solutions are, for the ``standard" initial conditions
$\langle\underline\sigma(t=0)\rangle=(0,0,z_0=1)^T$,
\begin{eqnarray}
\langle\sigma_x(s)\rangle&=&{1\over
s+{E^2\over\Delta^2}\Gamma_1(s)}\left(\left(\epsilon-{E\over\Delta}K_y^+(s)\right){N(s)\over
D(s)}-{E\over\Delta}k^-(s)\right),\label{p2_11e2_541}\\
\langle\sigma_y(s)\rangle&=&-{N(s)\over
D(s)},\label{p2_11e2_542}\\
\langle\sigma_z(s)\rangle&=&-{\Delta\over s}{N(s)\over
D(s)}+{z_0\over s}\,\,\, ,\label{p2_11e2_543}\\
N(s)&=&{E\over\Delta}\left(\epsilon-{E\over\Delta}K_y^+(s)\right)k^-(s)+\left({\Delta\over
s}z_0+k_y^-(s)\right)\left(s+{E^2\over\Delta^2}\Gamma_1(s)\right),\label{p2_11e2_544}\\
D(s)&=&\left(s+\Gamma_y(s)+{\Delta^2\over
s}\right)\left(s+{E^2\over\Delta^2}\Gamma_1(s)\right)+
\left(\epsilon-{E\over\Delta}K_y^+(s)\right)^2.\label{p2_11e2_545}
\end{eqnarray}
To go further, we need an explicit expression for the bath
correlator $C(t)$.  For the spin-boson model, the well-known
formula is
\begin{equation}
C(t)=\int_0^\infty d\omega
J(\omega)(\coth({\beta\omega/2})\cos(\omega t)+i\sin(\omega
t)).\label{p7_2e48}
\end{equation}
For the ohmic case, Eq. (\ref{ohm1}), Eq. (\ref{p7_2e48}) becomes
\begin{equation}
C(t)=-{\alpha\over\beta^2}{\rm
Re}~\psi'\left({1-i\omega_ct\over\beta\omega_c}\right)-{\alpha\omega_c^2\over
2(i+\omega_c t)^2}~~,\label{DDV25Oct2002}
\end{equation}
where $\psi'$ is the derivative of the digamma function\cite{AS}.
We are not aware that this simple exact formula has appeared
previously in the literature.

\section{Markovian limit}

For discussing the exact solution it is instructive to understand
the structure of the solution in a Markov approximation.  This
approximation is obtained by replacing all the kernels $\Gamma_1$,
$\Gamma_y$, $K_y^+$, $k^-$, and $k_y^-$ by their forms near $s=0$.
For all except $k^-$, this means replacing them by constants;
$k^-$ has a $1/s$ divergence at small $s$.  Then the solutions
Eqs. (\ref{p2_11e2_543}) are rational functions of $s$. If the
poles of these rational functions are located at positions $s_k$
in the complex $s$ plane, with residues $r_k/2\pi i$, then the
inverse Laplace transform can be written
$\langle\sigma_\mu(t)\rangle=\sum_k r_k^\mu \exp(s_i t)$. We
indicate here that while the residues do depend
on the label $\mu=x,y,z$, the pole positions do not, as is shown
by the form of Eqs. (\ref{p2_11e2_543}).

As is well known\cite{weiss}, there are four poles at positions
\begin{equation}
s_1=0,\,\,\,\,\,\, s_2=-\Gamma_1^0,\,\,\,\,\,\,
s_{3,4}=-\Gamma_2^0\pm i{\tilde E}.\label{polpos}
\end{equation}
  The first pole
describes the long-time asymptote of the solution (stationary
state), the second the purely exponential, ``$T_1$"-type decay
(relaxation), and the last two (complex conjugate paired) describe
an exponentially decaying sinusoidal part, the ``$T_2$"-type decay
of coherent oscillations.  The expressions for the constants in
Eq. (\ref{polpos}) are, to lowest order in $\alpha$, given by
\begin{eqnarray}
\Gamma_1^0&=&T_1^{-1}={\alpha\pi\Delta^2\over E}\coth(\beta
E/2),\\
\Gamma_2^0&=&T_2^{-1}={1\over
2}\Gamma_1^0+{2\alpha\pi\epsilon^2\over E^2}k_BT,
\label{p5_8e5_292}
\end{eqnarray}
and\cite{GS}
\begin{eqnarray}
{\tilde E}=E+\delta E&,&\,\,\,\,\,\delta E=\delta E^{\rm
Lamb}+\delta E^{\rm Stark},\\
\delta E^{\rm Lamb}&=&{\alpha\Delta^2\over E}\left (C -
\ln{\omega_c\over E}\right ),\label{lamb}\\
\delta E^{\rm Stark}&=&{\alpha\Delta^2\over E}({\rm
Re}~\psi(iE\beta/2\pi)-\ln(E\beta/2\pi)),
\end{eqnarray}
where we have dropped terms of order $E/\omega_c$ and higher, C is
the Euler constant, and $\psi$ is the digamma function \cite{AS}.
These expressions are straightforwardly derivable, and agree with
the literature\cite{weiss}, except for the energy shift due to
vacuum fluctuations, $\delta E^{\rm Lamb}$ (which contains in
general $\ln(\omega_c/E)$ and not $\ln(\omega_c/\Delta)$).

The residues of these poles are, in the limit $\alpha\rightarrow
0$,
\begin{equation}
\begin{array}{lll}
r_1^x=x_\infty=-(\Delta/E)~\tanh(\beta E/2),&
r_1^y=y_\infty=0,&r_1^z=z_\infty=-(\epsilon/E)~\tanh(\beta
E/2),\\
r_2^x=\epsilon\Delta/E^2-x_\infty,&r_2^y=0,&
r_2^z=\epsilon^2/E^2-z_\infty,\\r_{3,4}^x=-\epsilon\Delta/2E^2,&
r_{3,4}^y=-\Delta/2E,&r_{3,4}^z=\Delta^2/2E^2.\end{array}
\label{p5_8e5_29}
\end{equation}
We note that this Markovian
theory satisfies the expected fundamental relation\cite{Abragam}
\begin{equation}
\Gamma_2^0=\Gamma_1^0/2+(2\epsilon^2/E^2)\int_{-\infty}^\infty
dt\langle X(t)X\rangle_B\,\,\,\,\,\ {\mbox{(Korringa relation);}}
\end{equation}
also, to lowest order in $\alpha$, the asymptotic values of
$\langle\sigma_\mu(t\rightarrow\infty)\rangle$ go to the Boltzmann
equilibrium distribution of the system, e.g.,
$z_\infty=-(\epsilon/E)~ \tanh(\beta E/2)$, unlike in the popular
``non-interacting blip" approximation\cite{weiss}.

\section{Branch cuts at T=0}

We now return to the exact solution, examining it in detail at
vanishing temperature $T=0$.  In this case Eq.
(\ref{DDV25Oct2002}) becomes
\begin{equation}
C_{T=0}'(t)={\alpha\omega_c^2\over
2}{1-\omega_c^2t^2\over(1+\omega_c^2t^2)^2},\,\,\,\,\,\,\,C_{T=0}''(t)=\alpha\omega_c^2
{\omega_ct\over(1+\omega_c^2t^2)^2},
\end{equation}
and the Laplace transform of $C$ is
\begin{eqnarray}
C_{T=0}'(s)&=&{\alpha s/2}\left(-\cos\left({\tilde s}\right){\rm
Ci}\left({\tilde s}\right)- \sin\left({\tilde s}\right){\rm
si}\left({\tilde
s}\right)\right)\nonumber\\\label{DDVmathematicac1anycut}
C_{T=0}''(s)&=&-{i\alpha/2}\left(-\omega_c + s~\sin\left({\tilde
s}\right){\rm Ci}\left({\tilde s}\right) - s\cos\left({\tilde
s}\right){\rm si}\left({\tilde s}\right)\right)\,
,\label{DDVmathematicac1anycut2}
\end{eqnarray}
where ${\tilde s}=s/\omega_c$\cite{footnotewc}. There is an
important feature of this correlation function that makes the
Markov solution qualitatively incomplete: while the sine integral
${\rm si}(s)$ is an analytic function of $s$, the cosine integral
${\rm Ci}(s)$ behaves like $\ln(s)$ for $s\rightarrow 0$\cite{AS}.
This means that $C(s)$ is nonanalytic at $s=0$ --- it has a branch
point there. Thus, the exact solutions
$\langle\sigma_\mu(s)\rangle$ have extra analytic structure not
present in the Markov approximation, and the real-time dynamics
$\langle\sigma_\mu(t)\rangle$ has qualitatively different features
in addition to the pole contributions we have just discussed.

The $s=0$ branch point in $C(s)$ leads the kernels $\Gamma_1(s)$,
$K_y^+(s)$, and $k^-(s)$ to have branch points at $s=\pm iE$; the
kernels $\Gamma_y(s)$ and $k_y^-(s)$ have three branch points, at
$s=0$ and $s=\pm iE$.  Thus, the solutions to the GME
$\langle\sigma_{x,y,z}(s)\rangle$ also have three branch points in
the complex plane.  We find by numerical study that the exact
solutions still have four poles as before, which, for small
$\alpha$, have nearly (but not exactly) the same pole positions
and residues as in the Markov approximation.

\begin{figure}[htb]
\epsfxsize=15cm \epsffile{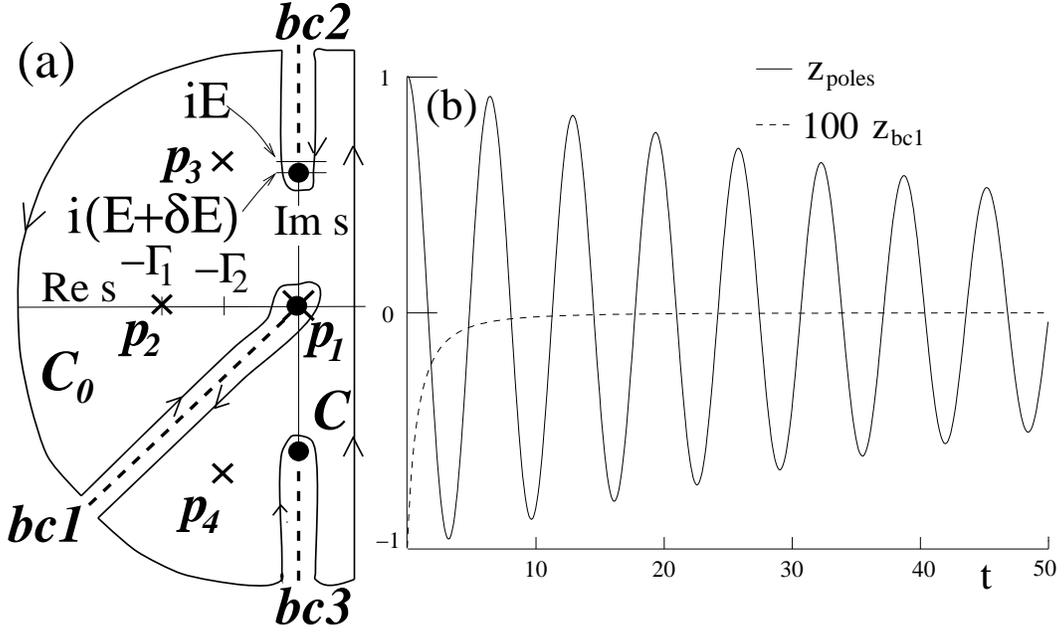} \caption{(a) Structure of
the solutions $\langle\sigma_\mu(s)\rangle$ in the complex $s$
plane.  The four poles $p_1$, $p_2$, $p_3$, and $p_4$, are
indicated by crosses; the three branch points at $s=0,\pm iE$ are
indicated by solid circles, and the three branch cuts chosen, bc1,
bc2, and bc3, are indicated by dashed lines.  The inverse Laplace
transform requires an integration along the contour ${\bf C}$
parallel to the imaginary axis.  This integral may be evaluated by
closing with a contour in the left half plane (${\bf C_0}$, the
Bromwich contour), which lies at infinity except for looping back
around each of the branch cuts. (b) $z_{poles}(t)$ and
$z_{bc1}(t)$ for the unbiased case, $\epsilon=0$, $\Delta=1$,
$\omega_c=30$, $T=0$, and $\alpha=0.01$.  $t$ is in units of $1/E$
(i.e., $E=1$).} \label{fig1}
\end{figure}

Thus, the structure of the solutions in the complex $s$ plane is
as shown in Fig. \ref{fig1}a.  The locations of the branch cuts are
chosen for computational convenience, as discussed shortly. Given
this branch-cut structure, the inverse Laplace transform (the
Bromwich integral) is evaluated by closing the contour as shown.
Thus, the exact inverse Laplace transform can be expressed as ($t>0$)
\begin{eqnarray}
\langle\sigma_\mu(t)\rangle&=&{1\over 2\pi i}\int_{\cal
C}dse^{st}\langle\sigma_\mu(s)\rangle={1\over 2\pi i}\oint_{{\cal
C}_o}dse^{st}\langle\sigma_\mu(s)\rangle\nonumber\\
&&-{1\over 2\pi i}\sum_{k=1}^3 q_k\int_{p_k}^{\infty}dxe^{q_kxt}
(\langle\sigma_\mu(q_kx+\eta_k)\rangle-
\langle\sigma_\mu(q_kx-\eta_k)\rangle). \label{DDV1_6_03}
\end{eqnarray}
Here $q_k=e^{i\theta_k}$ and $\eta_k=\eta e^{i(\theta_k-\pi/2)}$,
with $\eta$ an infinitesimal positive real number.  That is,
$\eta_k$ is an infinitesimal displacement perpendicular to the
direction of branch cut $k$.  For the cut choices we have made,
$\theta_1=5\pi/4$, $\theta_2=\pi/2$, $\theta_3=3\pi/2$, $p_1=0$,
and $p_2=p_3=E$. The closed-contour integral in the expression can
be written as a sum over the four poles, and so gives complex
exponential contributions to the solution as in the Markovian
case.  The extra terms, the sum over the three branch cuts, are
new and give qualitatively different features.  The contributions
of the second and third branch cuts are complex conjugates of each
other, so we will be discussing them together.

The contribution of these cuts to the solution is independent of
the detailed positioning of the branch cuts, so long as they are
not moved across a pole; the choice of the direction of bc1 is a
computational convenience --- the apparently most natural choice
of this cut direction, along the negative real axis, passes it
essentially on top of the $\Gamma_1$ pole, making the evaluation
of the branch-cut integral numerically inconvenient. As a check,
we find that the results we discuss now are indeed independent of
the cut direction.

We will do a detailed study of these branch-cut contributions for
$\langle\sigma_z(t)\rangle\equiv z(t)$.  We will use the following
notation for the branch cut terms in Eq. (\ref{DDV1_6_03}); for
``branch cut 1" (bc1),
\begin{equation}
z_{bc1}(t)= -{1\over 2\pi i}q_1\int_{p_1}^{\infty}dxe^{q_1xt}
(\langle\sigma_z(q_1x+\eta_1)\rangle-
\langle\sigma_z(q_1x-\eta_1)\rangle),
\end{equation}
and for two complex-conjugate cuts denoted together as ``branch
cut 2" (bc2):
\begin{equation}
z_{bc2}=-{1\over 2\pi i}\sum_{k=2}^3
q_k\int_{p_k}^{\infty}dxe^{q_kxt}
(\langle\sigma_z(q_kx+\eta_k)\rangle-
\langle\sigma_z(q_kx-\eta_k)\rangle).
\end{equation}

\subsection{Unbiassed case}

For the unbiased spin-boson case, $\epsilon=0$, an essentially
analytic calculation can be done for all contributions; we find
that these agree, as expected, with the weak-coupling limit of the
calculations presented in \cite{Leggett}.  In this case there is
no bc2 contribution, $z_{bc2}(t)=0$ for all $t$.  The bc1
contribution can be obtained analytically to leading order in
$\alpha$: $z(t)=z_{poles}(t)+z_{bc1}(t)$,
\begin{equation}
z_{bc1}(t)=-\alpha\{1-\Delta t[{\rm Ci}(\Delta t)\sin(\Delta
t)-{\rm si}(\Delta t)\cos(\Delta t)]\}.\label{p8_7e8_24}
\end{equation}
This function, plotted along with the pole contribution in Fig.
\ref{fig1}b for the choice of parameters shown, has the following
features: $z_{bc1}(t)$ is negative for all $t$, it is
monotonically increasing, and its long-time behavior is
$z_{bc1}(t)\sim-2\alpha/(\Delta t)^2$. Also,
$z_{bc1}(t=0)=-\alpha$.

Let us survey, then, the peculiar features that this branch cut
contribution introduces into the time response $z(t)$. Visualizing
the $\epsilon=0$ spin-boson model as a symmetric double well
system coupled to its environment, the bc1 piece being negative
means that, if the system is initially in the left well, it will,
in the course of coherently tunnelling back and forth, spend more
time in the {\em right} well!  This effect becomes strongest at
long time, much longer than $T_2$, for in this regime the pole
contributions are exponentially small, while the bc1 contribution
decays like a power law. Experimentally it may be hard to see the
effect in this regime (on account of finite-temperature effects,
for example), so it is important to note that this memory effect
appears already at early times, indicating that already in the
first couple of coherent oscillations, there will be an excess
amplitude in the right-well excursions as compared with the
left-well excursions, by an amount proportional to $\alpha$. We
judge, on the basis of a variety of evidence\cite{foot2}, that the
Born approximation should be reliable up to $\alpha$'s of order
$1-2\%$; thus, experiments that look at coherent oscillations
accurately at the percent level (which, it seems, will ultimately
be necessary for performing quantum computation) could readily see
this bc1 effect.

We note several other interesting features of our solution for
$\epsilon=0$.  Taking into account the non-Markovian effects, we
can do a more precise calculation of the pole positions and
residues (only poles 3 and 4 contribute). We find, for $T=0$,
$\Gamma_2\equiv -{\rm Re}(s_3)=\Gamma_2^0 r$, where, as before
$\Gamma_2^0=\alpha \pi\Delta/2$, and the renormalization factor
$r$ is given by $r =(1-\alpha)/(\kappa^2+\alpha^2\pi^2)<1$, with
$\kappa=1-2\alpha(1/2+C-\ln(\omega_c/\Delta))$. Further, ${\rm
Im}(s_3)=E+\delta E^{\rm Lamb}{\tilde r}$, with ${\tilde
r}=(\kappa-\alpha\pi^2/2(C-\ln(\omega_c/\Delta)))/(\kappa^2+(\alpha\pi)^2)$.
These expressions are obtained as systematic expansions in the
small parameters $\Gamma_2/E$ and $\delta E/E$, and they match a
direct numerical evaluation of the pole positions very well up to
$\alpha$'s of a few percent. For the corresponding pole residues
we find the simple result in leading order $r_3+r_4=1+\alpha
+O(\alpha^2)$.  This would be impossible in a Markovian theory, in
which $z(t=0)=r_3+r_4$, so that $r_3+r_4$ would be exactly 1 to
all orders in $\alpha$.  In fact this excess pole residue is
exactly what is needed to cancel out the initial value of the bc1
contribution to $z(t)$.
We note that our results for the residues differ from the
weak-coupling expressions in the literature\cite{weiss} (we are
not aware of prior reports on the renormalization factors $r$ and
${\tilde r}$).

\subsection{Biassed case}

For the biased model ($\epsilon\neq 0$) the bc2 contributions
become nonzero; we find that they give other peculiar
non-exponential corrections to the solution $z(t)$, very different
from the bc1 contribution.  The previous ``NIBA" calculations of
\cite{Leggett} are inapplicable in this case, and our results here
are completely new.  We can do a nearly analytic evaluation of the
bc2 contribution to Eq. (\ref{DDV1_6_03}):  Using Eq.
(\ref{p2_11e2_543}) and expanding to lowest order in $\alpha$, we
find for the integrand of the sum of the $k=2$ and 3 terms of
(\ref{DDV1_6_03}),
\begin{equation}
z_{bc2}(s=i\omega)\approx{2\Delta^2\over
\omega}{b^-(\omega)\over(E^2-\omega^2+b^+(\omega))^2+b^-(\omega)^2}.\label{p10_3e10_8}
\end{equation}
Here $b(i\omega \pm \eta)\equiv b^+(\omega)\pm ib^-(\omega)$,
$b(s)\equiv\alpha(d(s)+n(s)(s^2+E^2)/\Delta)$, where $d(s)$ and
$n(s)$ are given by $N(s)=\Delta+\alpha n(s)$ (see Eq.
(\ref{p2_11e2_544})) and $D(s)=s^2+E^2+\alpha d(s)$ (see Eq.
(\ref{p2_11e2_545})).  Since $b^-(\omega)=0$ for $|\omega|\leq E$,
it is reasonable to expect that $b^-$ will grow linearly as one
passes onto the branch cut; and, in fact, we find from numerical
study that a good ansatz is $b^-(\omega)=(E-\omega){\tilde
b}^-(\omega)$, with ${\tilde b}^-(\omega)$ being a weakly varying,
real function of $\omega/E$. With this, for $\omega$ of order $E$,
Eq. (\ref{p10_3e10_8}) simplifies to
\begin{equation}
z_{bc2}(s=i\omega)\approx-{\Delta^2{\tilde b}^-(E)\over 2E^3}{1\over
\omega-E}.\label{p10_4e10_13}
\end{equation}
We find that (\ref{p10_4e10_13}) should be valid for
$\omega>E+b^+(E)/2E$.  Using (\ref{p10_4e10_13}) we can do the branch
cut integral, which gives (for $t\leq 1/(\alpha E x_0)$ --- see
Appendix for an alternative approach),
\begin{equation}
z_{bc2}(t)\approx\alpha x_1\log(x_0 \alpha
Et)\cos(Et+\phi).\label{DDVbc2note4}
\end{equation}
Here $\phi$ is a constant phase shift that we have not determined
explicitly (but see Appendix), and the dimensionless constants
$x_0$ and $x_1$ are
\begin{eqnarray}
x_0&=&|b^+(E)|/2\alpha E^2=
|\delta E|/\alpha E\\
x_1&=&\Delta^2{\tilde b}^-(E)/2\alpha E^3. \label{p10.4e10.14-16}
\end{eqnarray}
Since $b^\pm\propto\alpha$, these constants are independent of
$\alpha$.  The last expression for $x_0$ comes from an evaluation
of $b^+(E)$: it is directly related to the energy renormalization
in the Markov approximation, $b^+(E)=2E\delta E^{\rm Lamb}$.

\begin{figure}[htb]
\epsfxsize=15cm \epsffile{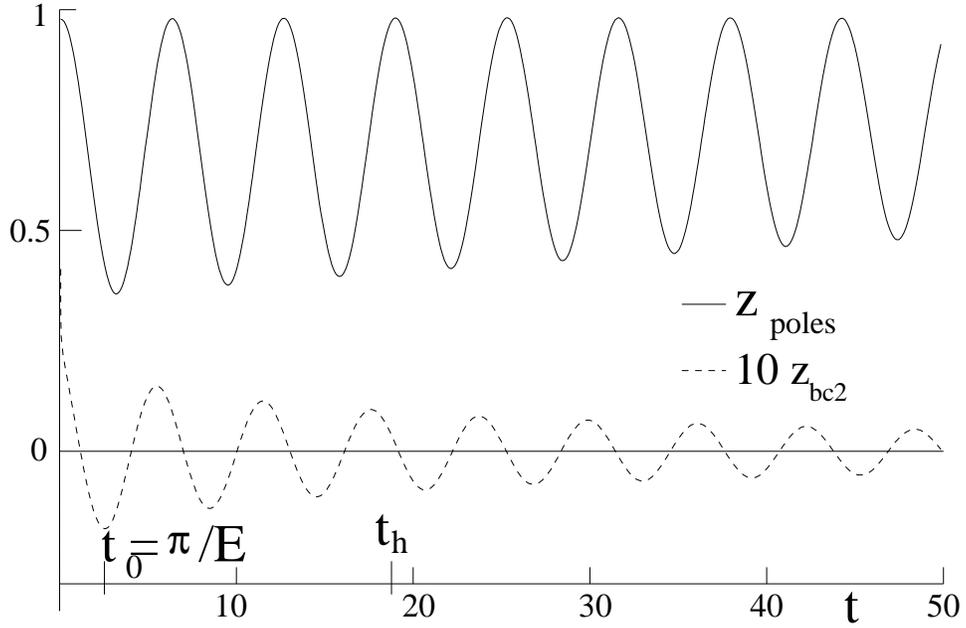} \caption{$z_{poles}(t)$ and
$z_{bc2}(t)$ for the biased case, illustrating the prompt loss of
coherence produced by bc2.  Here $E=1$,
$\epsilon/\Delta=-1.38$,
$\omega_c=30$,
$T=0$, and $\alpha=0.01$.  For these parameters, the time scale
for the prompt loss of coherence (using Eq.
(\protect\ref{p10_6e10_23})) is $t_h=18.98$.  $t_h$ is the time at
which the envelope of $z_{bc2}$ falls to half its value at
$t_0=\pi/E$.  This time scale is much shorter than the regular
exponential decay of coherence in $z_{poles}$; for our parameters,
$T_2=204.6$.} \label{fig2}
\end{figure}

In Fig.~\ref{fig2} we show a direct numerical evaluation of
$z_{bc2}(t)$. One can see the decay of the oscillatory part, which
is logarithmic according to Eq. (\ref{DDVbc2note4}).  Even though
the decay is very non-exponential, it is reasonable to attempt to
characterize this decay by a time scale.  Eq. (\ref{DDVbc2note4})
obviously does not work at $t=0$, since it is logarithmically
divergent. This is not surprising, since our calculation has
neglected cutoff effects (dependence on $\omega_c$), so Eq.
(\ref{DDVbc2note4}) is not expected to be correct for
$t<1/\omega_c$.  However, if we consider ``early" time to be the
first half-period of the coherent oscillation, $t_0=\pi/E$, then
Eq. (\ref{DDVbc2note4}) should be valid and we can use it to
characterize the decay by determining the time $t_h$ at which
$z_{bc2}(t)$ decreases to half its early-time value, i.e.,
$z_{bc2}(t_h)={1\over 2}z_{bc2}(t_0)$. We obtain
\begin{equation}
t_h={1\over  E}\sqrt{\pi E\over |\delta E|}\propto {1\over
E}{1\over\sqrt{\alpha}}\,.\label{p10_6e10_23}
\end{equation}
Surprisingly, $t_h\propto 1/\sqrt{\alpha}$ depends
non-analytically on $\alpha$. This explains the effect that is
evident in Fig. \ref{fig2}: for small $\alpha$, $t_h\ll T_2$, that
is, on the scale of $T_2$, there is a very rapid loss of coherence
as contributed by bc2.  This phenomenon may be called a {\em
prompt loss of coherence}, as it would appear experimentally as a
fast initial loss of coherence (from 100\% to $(1-c\alpha)100\%$,
$c$ being some constant near unity), followed by a much slower,
exponential decay of coherence on the regular $T_2$ time scale.

We make a few final remarks about the bc2 calculation.  The
absolute size of the bc2 contribution reaches a maximum near the
value of $\epsilon/\Delta$ used in Fig. \ref{fig2}; the relative
size of this contribution continues to increase as
$|\epsilon|/\Delta$ increases, so that it eventually becomes much
larger than the pole contribution (but all contributions to $z(t)$
go to zero as $|\epsilon|/\Delta\rightarrow\infty$).  When
$|\epsilon|\approx\Delta$, we find that, because of the prompt
loss of coherence, there is a {\em deficit} in the total pole
contribution, that is, $\sum_kr_k=1-O(\alpha)<1$.  Even in the
absence of an explicit branch cut computation, this deficit
signals the prompt loss of coherence, in that it indicates that
the exponentially decaying contributions to $z(t)$ do no account
for all the coherence near $t=0$.  Note that this is opposite to
the unbiased case, where, as a result of the bc1 part, there is an
{\em excess} pole contribution.

\section{Next order calculation: steady state solution}

Finally, we present the result of a calculation of the corrections
to order $\alpha$ to the steady state (long time) solution of the
GME.  To this order, as we will now show, the spin does not go to
the Gibbs distribution of the uncoupled system (i.e., at $T=0$,
the ground-state density matrix of the isolated spin).  However,
the result is consistent with the Gibbs distribution of the {\em
coupled} system, giving good evidence for a strong form of
ergodicity, even at $T=0$.

These apparently simple corrections, reported below, require an
enormous additional calculation, in that they require an
evaluation of the next order of the Born series.  That is, we must
take the expansion of the self energy superoperator $\Sigma$ to
fourth order in $L_{SB}$. The formal expression for $\Sigma^{(4)}$
is simple enough to generate: it is well known that the full Born
series is generated by repeated substitution of the following
propagator identity into the exact expression for $\Sigma$ (Eq.
3.4.7a of \cite{FS}):
\begin{equation}
e^{-iQLt}=e^{-iQL_0t}-i\int_0^tdt_1e^{-iQL(t-t_1)}QL_{SB}e^{-iQL_0t_1}.
\label{FickSauer3_4_7a}
\end{equation}
Here
\begin{equation}
L_0=L_S+L_B.
\end{equation}
This generates the superoperator expression for $\Sigma^{(4)}$:
\begin{equation}
\Sigma^{(4)}(t)=(-i)^3\int_0^t dt_1\int_0^{t_1} dt_2P{\rm
Tr}_BL_{SB}e^{-iQL_0(t-t_1)}QL_{SB}e^{-iQL_0(t_1-t_2)}QL_{SB}e^{-iQL_0t_2}L_{SB}\rho_B
\label{p7tutor0}
\end{equation}
This expression can be simplified with the use of the operator
identities
\begin{equation}
QL_0=L_0,\,\,\,\,\,PL_{SB}Q=PL_{SB}. \label{p6_eq31}
\end{equation}
Only one factor of the projection superoperator $Q=1-P$ survives:
\begin{equation}
\Sigma^{(4)}(t)=(-i)^3\int_0^t dt_1\int_0^{t_1} dt_2{\rm
Tr}_BL_{SB}e^{-iL_0(t-t_1)}L_{SB}e^{-iL_0(t_1-t_2)}QL_{SB}e^{-iL_0t_2}L_{SB}\rho_B.
\label{p7tutor1}
\end{equation}
Note also that the projector $P$ has been dropped from the
expression; since it is immediately followed by a trace over the
bath it acts as the identity.  We can also write $\Sigma^{(4)}$ in
several equivalent convenient forms using the identity
\begin{equation}
e^{-iL_0t}L_{SB}=L_{V(-t)}e^{-iL_0t},\,\,\,\,V(t)=e^{iL_0t}H_{SB}.
\label{p7tutor2}
\end{equation}
This gives the following two equivalent forms for the self energy:
\begin{equation}
\Sigma^{(4)}(t)=(-i)^3\int_0^t dt_1\int_0^{t_1} dt_2{\rm
Tr}_BL_{V(0)}L_{V(t_1-t)}QL_{V(t_2-t)}L_{V(-t)}\rho_Be^{-iL_St},
\label{p7tutor3}
\end{equation}
\begin{equation}
\Sigma^{(4)}(t)=(-i)^3\int_0^t dt_1\int_0^{t_1} dt_2e^{-iL_St}{\rm
Tr}_BL_{V(t)}L_{V(t_1)}QL_{V(t_2)}L_{V(0)}\rho_B. \label{p7tutor4}
\end{equation}
Equation (\ref{p7tutor3}) can be used to evaluate corrections to
the last term of Eq. (\ref{p2_2e2_61}); we must add to Eq.
(\ref{p2_2e2_63}) a term of the form
\begin{equation}
I_{\mu\nu}^{(4)}(t)={i\over 2}{\rm
Tr}_S\sigma_\mu\Sigma^{(4)}(t)\sigma_\nu.\label{p7tutor5}
\end{equation}
The bath part of these traces require the fourth order bath
correlator, which using Wick's theorem is, at $T=0$,
\begin{eqnarray}
{\rm
Tr}_B[X(t_1)X(t_2)X(t_3)X(t_4)\rho_B]={\alpha^2\omega_c^4\over
4}\times\,\,\,\,\,\,\,\,\,\,\,\,\,\,\,\,\,\,\,\,\,\,\,\nonumber\\
\left [ {1\over(i+\omega_c(t_3-t_2))^2(i+\omega_c(t_4-t_1))^2}
+{1\over(i+\omega_c(t_3-t_1))^2(i+\omega_c(t_4-t_2))^2}\right
.\nonumber\\ \left .
+{1\over(i+\omega_c(t_2-t_1))^2(i+\omega_c(t_4-t_3))^2}\right ].
\label{p7tutor6}
\end{eqnarray}
Eqs. (\ref{p7tutor3}-\ref{p7tutor6}) are the starting point of our
next-order calculation of the $s=0$ residue, which gives the
long-time asymptote of the density matrix.  Every detail of this
calculation is presented in the accompanying Mathematica notebook.
It can be understood why computer algebra is necessary for the
completion of this calculation if one considers the complexity of
the above expressions when written out in ordinary operator form.
The four nested commutators generated by the Liouvillian produces
thousands of distinct terms, which all need to be integrated and
studied in the limit of $\omega_c/E\rightarrow\infty$.

To illustrate the complexity of this calculation, we give in
Appendix B one example of a relatively ``simple" intermediate
result (the integral form for $I_{xx}^{(4)}(t)$) that is obtained
in the Mathematica notebook.

Given the enormity of the calculation, the final result is very
simple:
\begin{eqnarray}
x_\infty&=&-{\Delta\over E}+\alpha\left [-{\Delta^3\over
E^3}+\left (C-\ln{\omega_c\over E}\right )\left ({\Delta^3\over
E^3}-{2\Delta\over E}\right )\right ],\\
z_\infty&=&-{\epsilon\over E}+\alpha{\epsilon\Delta^2\over
E^3}\left (C-1-\ln{\omega_c\over E}\right ).
\end{eqnarray}
Recall that $y_\infty=0$ exactly in the spin-boson model.  In this
expression all terms that vanish in the limit of $\omega_c/
E\rightarrow\infty$ have been dropped.  Note that as in the
$\delta E$ calculation above, we see a mild (logarithmic)
divergence with the ultraviolet cutoff; all physical quantities
that we have calculated at this order have no divergence more
severe than this.  These results differ with the $O(\alpha)$ limit
results reported in Sec. 21.5.2 of \cite{weiss}; we can offer no
explanation for this.  There is no obvious way of treating the
logarithmic divergences in $x_\infty$ and $z_\infty$ by
introduction of a renormalized $\Delta$ and $\epsilon$, except in
the unbiassed case.  Nevertheless, the expressions given are
perfectly physical ($x_\infty^2+z_\infty^2<1$) within the expected
limits ($\omega_c>>E$, and $\alpha<1/\ln{\omega_c\over E}$).

After obtaining the above results, we separately calculated the
equilibrium density matrix, i.e.,
\begin{equation}
\langle\sigma_\mu\rangle_{eq.}={{\rm Tr}\sigma_\mu e^{-\beta
H}\over Z}=-{2\over \beta}{\partial\over\partial c_\mu}\ln Z,
\end{equation}
in the limit $T\rightarrow 0$ and for large $\omega_c$.  Here
$Z={\rm Tr} e^{-\beta H}$, $c_x=\Delta$, and $c_z=\epsilon$.  We
find that
\begin{equation}
\lim_{\beta\rightarrow \infty}{1\over\beta}\ln Z={1\over
2}(E+\delta E^{\rm Lamb}+\alpha\omega_c),
\end{equation}
with $\delta E^{\rm Lamb}$ from Eq. (\ref{lamb}).  Then it is a
simple calculation to show that the equilibrium and steady state
solutions actually coincide, i.e.,
\begin{equation}
x_\infty=\langle\sigma_x\rangle_{eq.},\,\,\,
z_\infty=\langle\sigma_z\rangle_{eq.}.
\end{equation}
While this result is natural, it should not be considered obvious;
it provides a rigorous demonstration that, to order $\alpha$, the
system is ergodic in a strong sense.

We give a few final notes about other quantities that require a
calculation of $\Sigma$ to the $\Sigma^{(4)}$ level. The
$O(\alpha)$ corrections to pole positions, given in an earlier
section, are unaffected by inclusion of $\Sigma^{(4)}$; however,
$O(\alpha)$ corrections to residues of both pole 1 ($s=0$) and
pole 2 for $\sigma_x$ and $\sigma_z$ are affected by
$\Sigma^{(4)}$.  $O(\alpha)$ corrections to $\sigma_y$ residues,
and $\sigma_{x,z}$ residues of poles 2 and poles 3 and 4 are
determined entirely by $\Sigma^{(2)}$; they do {\em not} have
contributions from $\Sigma^{(4)}$.


\section{Discussion}

Naturally, many more regimes could be studied using the present
approach.  For finite temperature the time evolution is very
different at long times, but it is essentially the same as the
$T=0$ evolution when $t<\hbar/kT$.  Recently, there has been
interest in varying both the system\cite{privman} and
bath\cite{karl} initial conditions, as well as in varying the
model of the bath density of states\cite{karl}.  For all these
circumstances, the systematic Born expansion procedure we report
here can be done.  It is clear on general grounds that the
appearance of branch cut contributions will not be restricted to
the ohmic model, however, the ohmic case is special in that the
size branch cut contribution is not governed by any small
parameter.  For any superohmic spectral density of the form
$J(\omega)\propto\omega^n$ at low frequencies ($n=1,2,...$),
$C(t)$ will have a power-law dependence at long time, and thus
$C(s)$ will have a branch point at $s=0$.  However, the magnitude
of the branch cut contribution in the general case goes like
$1/w_c^{n-1}$.  So, non-exponential contributions to the dynamics
vanish in the physical limit in all these other cases.

Our hope is that, using the present and further exact calculations
of the weak-coupling behavior of the spin-boson model, a tool will
be made available to permit precision experiments to test the
validity of the model (which, at present, is only
phenomenologically justified) in various physical situations of
present interest in quantum information.  A fundamentally correct,
experimentally verified theory of the system and its environment
should ultimately be of great value in finding a satisfactory
qubit for the construction of a quantum information processor.

\section*{Acknowledgements}

DPDV thanks Armin Allaverdyan and Theo Nieuwenhuizen for useful
discussions.  He is supported in part by the National Security
Agency and the Advanced Research and Development Activity through
Army Research Office contract number DAAD19-01-C-0056.  He also
thanks the Institute for Quantum Information at Cal Tech
(supported by the National Science Foundation under Grant. No.
EIA-0086038) for its hospitality during the initial stages of this
work, and the Institute for Theoretical Physics at the University
of Amsterdam for hospitality during the work's late stages.  DL
thanks the Swiss NSF, NCCR Nanoscience, DARPA and the ARO.

\appendix
\section{Scaling form for branch cut integrals}

By numerical study we find that the branch cut integrals conform
to some simple scaling laws for small $\alpha$.  If we write the
bc1 and bc2 integrals as $z_{bc1}(t)=\int_0^\infty dx e^{-q_1xt}
z_{bc1}(s=q_1x)$ and $z_{bc2}(t)={\rm Re}\int_E^\infty dx e^{ixt}
z_{bc2}(s=ix)$, then we find that for small $\alpha$ and for
$s<<\omega_c$, $z_{bc1}(s)$ can be written in a scaling form
\begin{equation}
z_{bc1}(x)=(\alpha/E)f_1(\epsilon/\Delta,x/E).\label{DDVbc2note1}
\end{equation}
But for bc2 a very different scaling law applies:
\begin{equation}
z_{bc2}(x)=(1/E)f_2(\epsilon/\Delta,(x-E)/\alpha
E).\label{DDVbc2note2}
\end{equation}
Here $f_{1,2}$ are dimensionless, ``universal" functions that
govern the behavior of the branch cut contributions for small
$\alpha$.  For bc1, the behavior that the scaling law gives is
very simple: Eq. (\ref{DDVbc2note1}) implies that
$z_{bc1}(t)=\alpha {\bar f}_1(\epsilon/\Delta,Et)$, where ${\bar
f}_1$ is the Laplace transform of the scaling function $f_1$.  We
might have expected this behavior from Eq. (\ref{p8_7e8_24}), from
which we can read off the scaling function for $\epsilon=0$.  In
fact it appears from numerical studies that $f_1$ hardly changes
as $\epsilon$ is varied, except for an overall scale factor; that
is, ${\bar f}_1(\epsilon/\Delta,Et)\approx
a(\epsilon/\Delta)b(Et)$.  We find that the scaling function
$a(\tau)>0$ is peaked at $\tau=0$.  So, the memory effect
described above for $\epsilon=0$ persists for finite $\epsilon$,
but becomes smaller. For $|\epsilon|\approx\Delta$ the bc2
contribution, which we will describe now, becomes dominant over
the bc1 one.

Returning to Eq. (\ref{DDVbc2note2}), if we write the Fourier
transform of the scaling function as ${\bar
f}_2(\tau)=\int_0^\infty e^{ix\tau}f_2(x)dx$ and consider its
polar form ${\bar f}_2(\tau)=r_2(\tau)e^{i\phi_2(\tau)}$, then we
obtain
\begin{equation}
z_{bc2}(t)=\alpha r_2(\alpha Et)\cos(Et+\phi_2(\alpha
Et)).\label{DDVbc2note3}
\end{equation}
This shows that bc2 contributes an oscillatory part to the
solution, whose ``$T_2$" decay is determined by the features of
the scaling function $r_2$.  A few more observations about $f_2$
(obtained initially from numerical study) reveal some crucial
properties of the $r_2$ function: 1) $f_2(0)=0$;  2) $|f_2(x)|$
has a single maximum at $x=x_0$, where $x_0$ is some constant of
order unity; 3) Most important for the present discussion, for
$x>x_0$ $f_2(x)$ approaches $1/x$, that is, $f_2(x)\sim x_1/x$,
where $x_1$ is another real constant of order unity.  Fact 3)
implies that, for $\tau\rightarrow 0$, $r_2(\tau)\approx
x_1\log(x_0\tau)$.  That is, we conclude that at sufficiently
short time (actually for $t\leq 1/(\alpha E x_0)$, so a relatively
long time),
\begin{equation}
z_{bc2}(t)=\alpha x_1\log(x_0 \alpha
Et)\cos(Et+\phi),\label{DDVbc2note4app}
\end{equation}
as stated in the text.

\section{$I_{xx}^{(4)}(t)$}

As an example of one of many, many intermediate results worked
through in the accompanying Mathematica notebook, we give here the
expression for $I_{xx}^{(4)}(t)$ (Eq. (\ref{p7tutor5})), in
``simplified" form:
\begin{eqnarray}
I_{xx}^{(4)}(t)=\int_0^tdt_1\int_0^{t_1}dt_2\left[-\frac{{\epsilon}^2\,\cos
(E\,t)}
     {E^2\,{\left( -t_1 -
           \frac{i }{\omega_c} \right) }^2\,
       {\left( t - t_2 -
           \frac{i }{\omega_c} \right) }^2}
  - \frac{{\epsilon}^2\,\cos (E\,t)}
   {E^2\,{\left( t_1 -
         \frac{i }{\omega_c} \right) }^2\,
     {\left( t - t_2 -
         \frac{i }{\omega_c} \right) }^2} - \right .\nonumber
         \end{eqnarray}
         \begin{eqnarray}
  \frac{{\epsilon}^2\,\cos (E\,t)}
   {E^2\,{\left( -t -
         \frac{i }{\omega_c} \right) }^2\,
     {\left( t_1 - t_2 -
         \frac{i }{\omega_c} \right) }^2} -
  \frac{{\epsilon}^2\,\cos (E\,t)}
   {E^2\,{\left( t - \frac{i }{\omega_c}
         \right) }^2\,{\left( t_1 -
         t_2 - \frac{i }{\omega_c}
         \right) }^2} -\nonumber
         \end{eqnarray}
         \begin{eqnarray}
  \frac{{\epsilon}^2\,\cos (E\,t)}
   {E^2\,{\left( -t_1 -
         \frac{i }{\omega_c} \right) }^2\,
     {\left( -t + t_2 -
         \frac{i }{\omega_c} \right) }^2} -
  \frac{{\epsilon}^2\,\cos (E\,t)}
   {E^2\,{\left( t_1 -
         \frac{i }{\omega_c} \right) }^2\,
     {\left( -t + t_2 -
         \frac{i }{\omega_c} \right) }^2} -\nonumber
         \end{eqnarray}
         \begin{eqnarray}
  \frac{{\epsilon}^2\,\cos (E\,t)}
   {E^2\,{\left( -t -
         \frac{i }{\omega_c} \right) }^2\,
     {\left( -t_1 + t_2 -
         \frac{i }{\omega_c} \right) }^2} -
  \frac{{\epsilon}^2\,\cos (E\,t)}
   {E^2\,{\left( t - \frac{i }{\omega_c}
         \right) }^2\,{\left( -t_1 +
         t_2 - \frac{i }{\omega_c}
         \right) }^2} -\nonumber
         \end{eqnarray}
         \begin{eqnarray}
  \frac{{\Delta}^2\,
     \cos (E\,t - E\,t_1 -
       E\,t_2)}{2\,E^2\,
     {\left( -t_1 -
         \frac{i }{\omega_c} \right) }^2\,
     {\left( t - t_2 -
         \frac{i }{\omega_c} \right) }^2} -
  \frac{{\Delta}^2\,
     \cos (E\,t - E\,t_1 -
       E\,t_2)}{2\,E^2\,
     {\left( t_1 -
         \frac{i }{\omega_c} \right) }^2\,
     {\left( t - t_2 -
         \frac{i }{\omega_c} \right) }^2} -\nonumber
         \end{eqnarray}
         \begin{eqnarray}
  \frac{{\Delta}^2\,
     \cos (E\,t - E\,t_1 -
       E\,t_2)}{2\,E^2\,
     {\left( -t - \frac{i }{\omega_c} \right)
         }^2\,{\left( t_1 -
         t_2 - \frac{i }{\omega_c}
         \right) }^2} -
  \frac{{\Delta}^2\,
     \cos (E\,t - E\,t_1 -
       E\,t_2)}{2\,E^2\,
     {\left( t - \frac{i }{\omega_c} \right)
         }^2\,{\left( t_1 -
         t_2 - \frac{i }{\omega_c}
         \right) }^2} -\nonumber
         \end{eqnarray}
         \begin{eqnarray}
  \frac{{\Delta}^2\,
     \cos (E\,t - E\,t_1 -
       E\,t_2)}{2\,E^2\,
     {\left( -t_1 -
         \frac{i }{\omega_c} \right) }^2\,
     {\left( -t + t_2 -
         \frac{i }{\omega_c} \right) }^2} -
  \frac{{\Delta}^2\,
     \cos (E\,t - E\,t_1 -
       E\,t_2)}{2\,E^2\,
     {\left( t_1 -
         \frac{i }{\omega_c} \right) }^2\,
     {\left( -t + t_2 -
         \frac{i }{\omega_c} \right) }^2} -\nonumber
         \end{eqnarray}
         \begin{eqnarray}
  \frac{{\Delta}^2\,
     \cos (E\,t - E\,t_1 -
       E\,t_2)}{2\,E^2\,
     {\left( -t - \frac{i }{\omega_c} \right)
         }^2\,{\left( -t_1 +
         t_2 - \frac{i }{\omega_c}
         \right) }^2} -
  \frac{{\Delta}^2\,
     \cos (E\,t - E\,t_1 -
       E\,t_2)}{2\,E^2\,
     {\left( t - \frac{i }{\omega_c} \right)
         }^2\,{\left( -t_1 +
         t_2 - \frac{i }{\omega_c}
         \right) }^2} -\nonumber
         \end{eqnarray}
         \begin{eqnarray}
  \frac{{\Delta}^2\,
     \cos (E\,t - E\,t_1 +
       E\,t_2)}{2\,E^2\,
     {\left( -t_1 -
         \frac{i }{\omega_c} \right) }^2\,
     {\left( t - t_2 -
         \frac{i }{\omega_c} \right) }^2} -
  \frac{{\Delta}^2\,
     \cos (E\,t - E\,t_1 +
       E\,t_2)}{2\,E^2\,
     {\left( t_1 -
         \frac{i }{\omega_c} \right) }^2\,
     {\left( t - t_2 -
         \frac{i }{\omega_c} \right) }^2} -\nonumber
         \end{eqnarray}
         \begin{eqnarray}
  \frac{{\Delta}^2\,
     \cos (E\,t - E\,t_1 +
       E\,t_2)}{2\,E^2\,
     {\left( -t - \frac{i }{\omega_c} \right)
         }^2\,{\left( t_1 -
         t_2 - \frac{i }{\omega_c}
         \right) }^2} -
  \frac{{\Delta}^2\,
     \cos (E\,t - E\,t_1 +
       E\,t_2)}{2\,E^2\,
     {\left( t - \frac{i }{\omega_c} \right)
         }^2\,{\left( t_1 -
         t_2 - \frac{i }{\omega_c}
         \right) }^2} -\nonumber
         \end{eqnarray}
         \begin{eqnarray}
  \frac{{\Delta}^2\,
     \cos (E\,t - E\,t_1 +
       E\,t_2)}{2\,E^2\,
     {\left( -t_1 -
         \frac{i }{\omega_c} \right) }^2\,
     {\left( -t + t_2 -
         \frac{i }{\omega_c} \right) }^2} -
  \frac{{\Delta}^2\,
     \cos (E\,t - E\,t_1 +
       E\,t_2)}{2\,E^2\,
     {\left( t_1 -
         \frac{i }{\omega_c} \right) }^2\,
     {\left( -t + t_2 -
         \frac{i }{\omega_c} \right) }^2} -\nonumber
         \end{eqnarray}
         \begin{eqnarray}
  \left . \frac{{\Delta}^2\,
     \cos (E\,t - E\,t_1 +
       E\,t_2)}{2\,E^2\,
     {\left( -t - \frac{i }{\omega_c} \right)
         }^2\,{\left( -t_1 +
         t_2 - \frac{i }{\omega_c}
         \right) }^2} -
  \frac{{\Delta}^2\,
     \cos (E\,t - E\,t_1 +
       E\,t_2)}{2\,E^2\,
     {\left( t - \frac{i }{\omega_c} \right)
         }^2\,{\left( -t_1 +
         t_2 - \frac{i }{\omega_c}
         \right) }^2}\right ].
\end{eqnarray}
This double integral, and many others, are fully evaluated in the
Mathematica notebook, in the large $\omega_c$ limit.

\end{document}